# Factors Influencing the Usage of Mobile Banking Apps Among Malaysian Consumers

Siti Nurdianah binti Mohamad Jalani and Sathishkumar Veerappampalayam Easwaramoorthy
School of Engineering and Technology, Sunway University, No. 5, Jalan Universiti, Bandar Sunway, 47500, Selangor Darul Ehsan, Malaysia
(Email: sathishv@sunway.edu.my)

## Abstract

Mobile banking apps have transformed the banking sector by offering customers with convenient, secure and easily accessible financial services. Even so, it is crucial for banks and the mobile banking apps developers to understand the factors that influence the utilisation of these apps among Malaysian consumer. This study will examine the influence of several factors which are security concerns, service quality, technological factors and convenience, on the usage of mobile banking apps. The study aims to discover the key factors that affect the usage of mobile banking apps. A quantitative research method was utilised, which involves the collection of data from an online survey. The survey managed to collect data from 152 respondents who are above 18 years old and users of mobile banking apps in Malaysia. The data was analysed with correlation analyses to examine the relationship between the variables. A multinominal logistic regression model was used as a predictive model to predict the usage of mobile banking apps. This study contributes to existing researches by highlighting the importance of security and convenience into the development and marketing strategies of mobile banking apps. The study can help them conduct improvements on their current apps and thus increase the usage of mobile banking apps among consumers in Malaysia.

# 1    Introduction

The financial services industry has been transformed by the recent advancements in mobile banking services, particularly the development of mobile banking apps. These apps provide consumers with a diverse array of services, including the management of personal financial accounts, making online transactions, and paying bills, just at the end of their fingertips. It offers consumers the efficacy of conducting financial transactions at their own convenience, without the necessity of physically visiting a bank. Nevertheless, the adoption and usage of mobile banking apps in Malaysia continue to differ significantly among Malaysians, despite the benefits and perks they offer. It is crucial for financial services institutions to comprehend the factors that influence the utilisation of these mobile banking apps, as it can assist in the enhancement of their own digital services, thereby allowing them to differentiate themselves from others in the financial service industry. The main goal of this study is to identify and analyse the factors that influence the usage of mobile banking apps. The objectives of the study are:

- To study the relationship between security concerns with the utilisation of mobile banking apps
- To study the effect of service quality on the utilisation of mobile banking apps
- To assess the effect of technological factors on the usage of mobile banking apps
- To examine the influence of convenience on the usage of mobile banking apps
- To predict the mobile banking apps usage among consumers in Malaysia

The financial service industry has undergone a transformation as a result of the growth of mobile banking services, particularly with the development of mobile banking apps. In spite of this technological advancement, some consumers continued to underutilise these mobile banking apps for their day-to-day financial activities. It is imperative for financial service institutions to comprehend the factors that contribute to this occurrence. This study will examine the factors that influence the usage of mobile banking apps, including security concerns, service quality, technological factors, and convenience. In turn, the study will provide banks with actionable insights to enhance their mobile banking services and increase the usage of the apps among Malaysian consumers. Majority of recent studies lack a comprehensive analysis of these factors within a single study, with most of the studies primarily examining mobile banking services as a whole, rather than focusing on mobile banking apps. Addressing this gap is crucial for banks in order for them to develop strategic measures that will increase the usage of mobile banking apps.

The study aims to answer the following research questions:

1. Do security concerns, service quality, technological factors, and convenience have any significant effect on the usage of mobile banking apps?
2. What is the relationship between security concerns and the usage of mobile banking apps?
3. Will service quality affect the usage of mobile banking apps?
4. Do technological factors have any significant effect on the usage of mobile banking apps?
5. What is the effect of convenience on the usage of mobile banking apps?
6. Can logistic regression model effectively predict the usage of mobile banking apps?

# 2    Literature review

The banking and financial sector has undergone a significant transformation because of the growth in digitalisation. This has led conventional financial institutions to transition by expanding their services online and with the continuous advancement in technology, banks have since developed their own mobile banking apps. These apps can provide consumers with the convenience and flexibility to manage their financial activities in real time without the necessity of physically visiting the bank. Despite this, there are still individuals who are sceptical and hesitant to fully utilise these apps. The factors that may influence this occurrence will be studied in the literature review. It will analyse factors such as security concerns, technological factors, convenience, and service quality and their effect on the usage of mobile banking apps. The objective of the study is to determine the relationship between these factors and the usage of mobile banking apps.

Mobile banking is a digital platform developed by banks and financial institutions to allow consumers to directly interact and utilises the services and products provided by the banks and financial institutions (Karim et al., 2020). Mobile banking is a form of Internet banking that allows consumers to access a bank's products and services through a secure website using smart devices such as tablets and smartphones, rather than accessing them through a desktop or a computer. Mobile banking apps on the other hand, are a form of software applications that are downloadable and can be operated on the smart devices (Tarawneh et al., 2021).

The surge in the usage of mobile banking has been driven by the rapid advancement in technology and the change in consumer preferences. In Malaysia, the number of subscribers that subscribe to mobile banking has increased from significantly since it was first introduced in 2005. According to the statistics by the Central Bank of Malaysia (BNM), the number of mobile banking subscribers in 2023 is 32.37 million, which is a significant increase from 17.23 million subscribers in 2019 (Bank Negara Malaysia, 2024). The statistics also showed that the mobile banking adoption rate among the population in Malaysia had reached 97% by the end of 2023.

One of the critical factors in the use of mobile banking apps is security concerns. This factor can greatly influence consumer's behaviour regarding the usage of mobile banking apps. Most people are hesitant to use mobile banking apps because of the fear on several possibilities such as potential data leaks and breaches, identity theft, and the general lack of trust or confidence in the safety of digital or online transactions. A study conducted by Kala Kamdjoug et al., (2021) demonstrated that consumers' intention to utilise mobile banking apps is heavily influenced by the concerns on the security and safety of their personal information. The study indicates that consumers are extremely sensitive on the security and privacy concerns associated with mobile banking apps. Thus is because since these apps are deployed and utilised on their personal devices, the risk of having their personal data leaked is higher. The study suggested that the outcome of their research can help banks or financial institutions in implementing effective strategies to protect and preserve consumers' data, hence promoting and ensuring the security of mobile banking apps.

The level of trust that users have in the privacy and security measures of a platform has a significant influence on their usage of mobile banking apps (Shahid et al., 2022). The study by Shahid et al., (2022) suggested that banks should prioritise the development of consumer confidence by ensuring the existence of a secure and dependable infrastructure of mobile banking apps. Singh & Srivastava, (2020) found that consumers' intention to utilise mobile banking apps are dependent on their perceptions of security. The study suggests that in order to ensure that consumers have confidence in the products and services offered, banks and financial institutions should employ a rigorous security measures in their mobile banking platforms.

Merhi et al., (2019) studied the factors that influence the adoption and utilisation of mobile banking services in England and Lebanon. The study revealed that the intention to utilise mobile banking services in both countries are significantly influenced by the perception of security and privacy. In order to ensure the highest level of security and safety, the researchers suggest that banks improve their IT infrastructures and mobile banking apps. The study also recommends that banks should enforce more stringent regulations to safeguard consumers' privacy and reduce the likelihood of personal data and information breaches. Purwanto et al., (2020) discovered a robust correlation between the level of trust users have in mobile banking apps and the security of the apps. The research posits that the security of an app is a critical factor in establishing the trust of consumers, as these apps utilise and retain their personal and financial information.

Balapour et al., (2020) conducted a study to investigate the influence of privacy-related perceptions on the security perceptions of mobile app users. In order to analyse this relationship, they implemented communication privacy management theory (CPM). The research revealed that the perceived level of security of mobile apps is adversely affected by the perceived degree of privacy risk. Conversely, the perceived efficacy of a privacy policy has a positive effect on the perceived level of security of mobile apps. Based on the level of sensitivity of the information managed by mobile banking apps, the study's results indicate that users' perspectives on privacy and security vary.

The adoption of mobile banking apps is significantly influenced by technological factors. Mobile banking apps are being enhanced with new features and improved user interfaces to optimise their functionality and performance as technology continues to evolve. These apps are becoming more innovative, utilising biometric authentication to improve security and AI-powered chatbots to expedite customer service. According to a study conducted by Giovanis et al., (2019), the adoption of mobile banking by consumers is influenced by technology-related factors, including the ease of use and performance expectancy. The research posits that in order to establish a competitive edge that sets them apart from their competitors, banks should establish a robust user-centric platform. Salem et al., (2019) discovered that users' intention to utilise mobile banking services is significantly influenced by technological factors. It is recommended that banks prioritise the improvement of their mobile app interface in order to gain a competitive edge over their competitors.

Contrary to this, a study conducted by El-Gohary et al., (2021) revealed that consumers have a diminished understanding of artificial intelligence technology. This result may be attributed to the limited adoption of technologies, such as chatbots, during the period of the study. At the time, AI-powered technology was regarded as relatively fresh. Nevertheless, it is expected that its popularity will continue to grow and it will be widely used. Hai Sam et al., (2023) discovered a robust correlation between user preference for mobile banking apps and technological advancements. The study suggests that consumers' intentions to utilise mobile banking apps are significantly influenced by those that provide secure and efficient transactions. The study propose that banks could improve the security of their mobile banking apps and provide supplementary features to attract a broader consumer base.

The significance of convenience on consumers' intentions to utilise mobile banking apps is paramount. This element plays a significant role in shaping users' preferences, and it can be considered a critical factor in determining their propensity to adopt mobile banking apps (Hai Sam et al., 2023).

In their 2020 study, Shankar & Rishi, studied the influence of various forms of convenience on the intention of users to utilise mobile banking. Consumer's intention to implement mobile banking is influenced by the ease and accessibility of Internet services, according to the study. In addition, the investigation determined that the simplicity of accessibility is an additional critical factor in determining an individual's intention to utilise mobile banking apps. The potential hazards of unauthorised disclosure of sensitive financial and personal information are well-known in the context of mobile banking app usage. However, Pal et al., (2021) discovered that the perceived convenience of mobile apps significantly influences a user's intention to use them. The risks that are associated with these apps are significantly outweighed by the extensive array of conveniences they provide. For the sustainability of a mobile banking app, convenience is one of the critical determinants, as per Shankar et al., (2022). The reason for this is that mobile banking apps allow users to access financial services from any location and at any time and are able to conduct transactions at a brisk pace without any difficulty.

In contrast, Jebarajakirthy & Shankar, (2021) discovered that the intention to utilise mobile banking apps was not influenced by convenience. This is due to the fact that the majority of banks provide the same products and services, which means that users' intention to implement mobile banking apps is not influenced by convenience. The study conducted by Nguyen, (2020) corroborates this conclusion, demonstrating that a user's intention to implement mobile banking is not significantly influenced by convenience.The intention of users to implement mobile banking applications is significantly influenced by the quality of service. Customers' satisfaction levels and their trust and loyalty can be enhanced by mobile banking apps that provide exceptional service. According to Hai Sam et al., (2023), there is a robust correlation between the intention of service quality consumers to utilise mobile banking applications in Malaysia. It is an essential factor, as numerous banks provide comparable services. Therefore, the quality of their service will significantly influence the intention of users to utilise mobile baking applications.

Khatab et al., (2019) evaluate the correlation between customer satisfaction levels and various aspects of service quality in institutions situated in Kurdistan. The study revealed that the level of customer satisfaction is significantly influenced by all of the parameters explored. It is crucial to maintain a consistently high level of customer satisfaction, as this will help to establish trust and loyalty among consumers towards institutions. This conclusion is supported by the study conducted by Li et al., (2021). Furthermore, the study found that consumers' satisfaction levels are greatly determined by the quality of services they received.

## 2.1 Frameworks

The conceptual framework for this study outlines the relationship between the independent variables: security concerns, service quality, technological factors, and convenience, with the dependent variable, the utilization of mobile banking applications. This framework functions as a visual depiction of the hypotheses and the direction of the study. In this framework:

- Security concerns: The level of security that mobile banking applications are believed to have
- Service Quality: The overall standard of services offered by the mobile banking applications
- Technological: The mobile banking apps' usability, functionality, and innovation
- Convenience: The convenience and accessibility of utilising mobile banking applications
- Mobile Banking Apps Adoption: The frequency at which users utilise mobile banking applications

Each independent variable is hypothesized to have a significant influence on the utilisation of mobile banking applications. This conceptual framework provides a foundation for studying the relationship between the variables in the study.

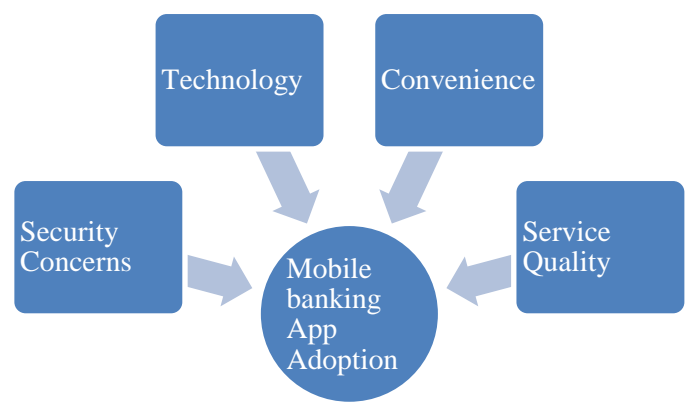

Figure 1: Conceptual Framework

This study's theoretical framework on the other hand are found on a number of well-established theories that elucidate consumer behaviour and the adoption of technology. These theories establish a fundamental foundation for understanding the factors that influence the utilisation of mobile banking applications.

Davis, (1989) introduced the Technology Acceptance Model (TAM), which can be used to explain the adoption of mobile banking applications among consumers. The theory posits that the decision to employ a technology is influenced by two factors that influence an individual's attitude towards the technology: the perceived utility and the perceived simplicity of use (Shachak et al., 2019). A limited number of studies implement the TAM model. One such study is that conducted by Al-Gharaibah, (2022), which employs the TAM model to examine the impact of perceived usefulness, customer attitude, convenience of use, and perceived risk on the adoption of e-banking in Malaysia.

Venkatesh et al., (2003) developed the Unified Theory of Acceptance and Use of Technology (UTAUT). In order to explain the behaviour associated with the adoption of technology, the model integrates a variety of factors, such as performance expectancy, effort expectancy, social influence, and facilitating environments. Tarawneh et al., (2021) employed the extended version of the Unified Theory of Acceptance and Use of Technology Model (UTAUT2) and the Model of Perceived Risk to investigate the adoption and utilisation of mobile banking among Generation Y in Malaysia.

The Decomposed Theory of Planned Behaviour (DTPB) and Diffusion of Innovations (DOI) can also be used to explain the adoption of mobile banking applications. Taylor & Todd, (1995) introduced the Decomposed Theory of Planned Behaviour, which deconstructs the components of Ajzen, (1991) Theory of Planned Behaviour (TPB). The Theory of Planned Behaviour posits that the behavioural intention is influenced by attitude, subjective norms, and perceived behavioural control, which in turn affects the actual behaviour. However, Rogers et al., (2014) Diffusion of Innovations posits that an individual's intention to adopt a technology or innovation is influenced by a variety of factors. Numerous studies incorporate these hypotheses into their research. For instance, Ho et al., (2020) conducted a study that integrates the Diffusion of Innovations (DOI), The Technology Acceptance Model (TAM), and the Decomposed Theory of Planned Behaviour (DTPB) to examine the factors that influence the intention to implement mobile banking. This study plans to offer a comprehensive comprehension of the factors that influence the use of mobile banking applications by incorporating these theoretical perspectives.

The existing studies have a few gaps and limitations, as indicated by the literature review. Initially, the majority of the studies concentrate solely on a particular country or region. Additional research is required to provide comparative comparisons across a variety of geographical regions. While research conducted in specific countries offers valuable insights into the local factors that influence the adoption of mobile banking applications, conducting a broader comparative research could illuminate the ways in which various aspects of culture and economy could influence the adoption rate of mobile banking applications worldwide..

Additionally, the majority of the study comprises a limited number of samples. This may result in potential bias and limiting the generalisability of their findings. A study conducted on a larger scale could offer more comprehensive insights into the factors that influence the adoption of mobile banking. Additionally, theoretical frameworks such as the TAM and UTAUT are employed in the majority of the investigations. The integration of these frameworks could offer a more comprehensive comprehension of the factors that influence the adoption of mobile banking. In order to enhance the relevance and efficacy of the research findings, it is crucial to resolve these gaps and constraints. By addressing these issues, it is possible to create more effective policies and practices that will promote the global adoption and usage of mobile banking apps.

# 3    Methodology

This study aims to explore the factors influencing the adoption of mobile banking apps among Malaysian consumers. The methodology employed in this research is carefully designed to address the research objectives through a combination of quantitative data collection and rigorous statistical analysis. The approach was selected to ensure that the findings are both reliable and relevant to the research questions.

The study was designed as a cross-sectional survey, which was appropriate for capturing data at a single point in time to assess the impact of security concerns, service quality, technological factors, and convenience on the frequency of mobile banking app usage. This design was chosen to efficiently access the effect of multiple factors on mobile banking apps usage.

A structured questionnaire was developed, consisting questions on demographic profiles and questions related to the key factors that influenced the usage of mobile banking apps. The questionnaire included:

- Demographic profile: Age, gender, education level, employment status, monthly income, and the state of residence.
- Mobile banking app usage: Questions on the frequency of mobile banking app usage, the mobile banking apps that currently being used, and the features of mobile banking apps
- Key Factors: A series of Likert-scale questions to measure security concerns, service quality, technological factors, and convenience.

## 3.1  Data collection

The sampling method used for the study is convenience method, where the survey was distributed online through social media platforms and through blasting to family and friends via messaging apps. The online distribution allows for a broad reach across different regions and demographic. A total of 152 responses were collected.

## 3.2  Data preparation

Before proceeding with data analysis, the data was cleaned and prepared by handling missing values. The missing values were imputed by using mode imputation. This was done to retain the size of the data since the data collected were relatively small. The variables were then encoded to transform the data. The independent variables were retained as continuous variables after averaging the Likert-scale responses.

## 3.3  Data analysis

There are several analyses conducted in the study:

- Spearman and Pearson correlation analysis were performed to examine the relationship between independent variables (Security Concerns, Service Quality, Technological Factors, and Convenience) with the dependent variable (Usage of Mobile Banking Apps). The correlation analyses will help to determine whether the factors have any influence on the usage of mobile banking apps.
- A multinomial logistic regression model was employed to predict the usage of mobile banking apps based on the key factors. The model's performance was evaluated using accuracy metrics, a confusion matrix, and a classification report, which helps to provide a comprehensive understanding of the predictive model.

### 3.4 Interpretation of results

The regression coefficients and p-values were analysed to determine the significance of each independent variables on the usage of mobile banking apps. The results are then interpreted to have a better understanding on which factors have the most influence in determining the usage of mobile banking apps among consumers in Malaysia.

### 3.5 Limitations

Several limitations were considered in the chosen methodology.

- The sampling method used for this study was convenience sampling method. The method was chosen because of the time constraint to conduct the study. Convenience sampling method, is a method used for the researcher to approach individuals who are most readily available to the researcher and who may be able to provide the necessary information (Mweshi & Sakyi, 2020). This method helps the researcher to collect the data in a short amount of time, however, since the link was shared to the people closest to the researcher, potential bias on the finding of the study may occurred since the possibility for the respondents to have similar demographic profiles and interests are high.
- The findings of the study cannot be used to represent the whole population. While online surveys allow the study to reach a broader population, there is a possibility that certain areas will be excluded unintentionally. such as those with limited internet access. This can produce a potential bias in the findings of the study.

### 3.6 Ethical considerations

The study was conducted with upholding the ethical research practices. The respondents were informed on the purpose of the study, the usage of their personal information in the study, and were also informed that their participation in the survey was not compulsory. The respondents were assured of the confidentiality and anonymity of their responses. They were also informed that the data collected from them will be used for the sole purpose of the study and will not be shared elsewhere. The study was conducted in compliance with the ethical guidelines, to ensure that the research was carried out with integrity and respect for the respondents.

## 4 Results and Discussion

### 4.1 Demographic profile

Table 1: Demographic Profile

| | Frequency | Percentage (%) | Valid Percentage (%) |
|---|---|---|---|
| **Gender** | | | |
| Male | 39 | 25.66 | 25.66 |
| Female | 113 | 74.34 | 100.00 |
| **Age** | | | |
| 18 – 20 | 2 | 1.32 | 1.32 |
| 21 – 30 | 99 | 65.13 | 66.45 |
| 31 – 40 | 22 | 14.47 | 80.92 |
| 41 – 50 | 20 | 13.16 | 94.08 |
| Above 50 | 9 | 5.92 | 100.00 |
| **Education Level** | | | |
| High school or equivalent | 2 | 1.32 | 1.32 |
| Diploma/Certificate | 18 | 11.84 | 13.16 |
| Bachelor's Degree | 101 | 66.45 | 79.61 |
| Master's Degree | 28 | 18.42 | 98.03 |
| Doctorate | 3 | 1.97 | 100.00 |
| **Monthly Income** | | | |
| Below RM 2000 | 21 | 13.82 | 13.82 |
| RM 2000 – RM 3999 | 47 | 30.92 | 44.74 |
| RM 4000 – RM 5999 | 42 | 27.63 | 72.37 |
| RM 6000 – RM 7999 | 20 | 13.16 | 85.53 |
| RM 8000 – RM 9999 | 9 | 5.92 | 91.45 |
| RM 10 000 and above | 13 | 8.55 | 100.00 |
| **State** | | | |
| Johor | 17 | 11.18 | 11.18 |
| Kedah | 1 | 0.66 | 11.84 |
| Kuala Lumpur | 34 | 22.37 | 34.21 |
| Melaka | 19 | 12.50 | 46.71 |
| Negeri Sembilan | 7 | 4.61 | 51.32 |
| Pahang | 2 | 1.32 | 52.63 |
| Perak | 6 | 3.95 | 56.58 |
| Pulau Pinang | 4 | 2.63 | 59.21 |
| Sabah | 2 | 1.32 | 60.53 |
| Sarawak | 1 | 0.66 | 61.18 |
| Selangor | 57 | 37.50 | 98.68 |
| Terengganu | 2 | 1.32 | 100.00 |

The Demographic Profile of the study is presented in Table 1. The respondents consist of mostly Female, which accumulate over 70% (n=113). Most of the respondents come from the age group between 21 to 30 years old which accumulate over 60% (n=99) of the data. Most of the participants has Bachelor's Degree (66.45%, n=101) as their highest level of education. The respondents also consists of those that earns between RM 2000 to RM 5999 (58.55%, n=89) every month with 30.92% (n=47) of them earns between RM 2000 – RM 3999, and the other 27.63% (n=42) earns between RM 4000 to RM 5999. Most of the respondents are staying in Kuala Lumpur and Selangor with 22.37% (n=34) from Kuala Lumpur and 37.50% (n=57) from Selangor.

### 4.2 Mobile Banking Apps Details
Table 2: Mobile Banking Apps Details

| **Most frequently used banking app** | | | |
|---|---|---|---|
| | Frequency | Percentage (%) | Valid Percentage (%) |
| CIMB OCTO (CIMB Bank) | 83 | 26.86 | 26.86 |
| GO (Bank Islam) | 37 | 11.97 | 38.83 |
| MAE (Maybank) | 104 | 33.66 | 72.49 |
| RHB Mobile Banking (RHB Bank) | 16 | 5.18 | 77.67 |
| Others | 69 | 22.33 | 100.00 |
| **Most used features** | | | |
| Checking out balance | 121 | 24.06 | 24.06 |
| Managing investments | 22 | 4.37 | 28.43 |
| Online transactions | 137 | 27.24 | 55.67 |
| Paying bills | 91 | 18.09 | 73.76 |
| Transferring money | 132 | 26.24 | 100.00 |
| **Frequency of banking app usage** | | | |
| Daily | 110 | 72.37 | 72.37 |
| Weekly | 35 | 23.03 | 95.39 |
| Monthly | 7 | 4.61 | 100.00 |

The usage details of the mobile banking apps are presented in Table 2. The two most popular mobile banking apps are MAE by Maybank and CIMB OCTO by CIMB Bank with 33.66% (n=104) and 26.86% (n=83) respectively. Most of the respondents use the mobile banking apps to check their account balance (24.96%, n=121), to conduct online transactions (27.24%, n=137), and transferring money (26,24%, n=132). Majority of the respondents use the mobile banking apps on a daily basis (72.37%, n=110).

### 4.3 Descriptive Analysis

Table 3: Descriptive analysis of factors influencing mobile banking app usage

| | *Security Concerns* | *Service Quality* | *Technological Factors* | *Convenience* |
|---|---|---|---|---|
| *Count* | 152 | 152 | 152 | 152 |
| *Mean* | 4.00 | 3.78 | 4.21 | 4.47 |
| *Standard Deviation* | 0.64 | 0.59 | 0.51 | 0.52 |
| *Minimum value* | 2.67 | 2.00 | 3.00 | 3.00 |
| *Maximum Value* | 5.00 | 5.00 | 5.00 | 5.00 |

The descriptive analysis in Table 3 provides an understanding oh how respondents perceive various factors that influence the usage of mobile banking apps. The relatively high means across all factors suggest that respondents generally have positive perceptions on the security, quality of service, technological factors, and convenience. Although most respondents are satisfied with the service quality of mobile banking apps they used, it still has a minimum value of 2 which indicates that a small group of the respondents have some dissatisfaction with the quality of services provided. With the relatively high average scores, most respondents are satisfied with the technological aspects of mobile banking apps they used and the convenience these apps provide them. As for security concerns, the standard deviation of 0.64 shows that the variable have a moderate spread of responses, indicating that there are some variation in the level of concern on the security of mobile banking apps.

**4.4 Correlation Analysis**

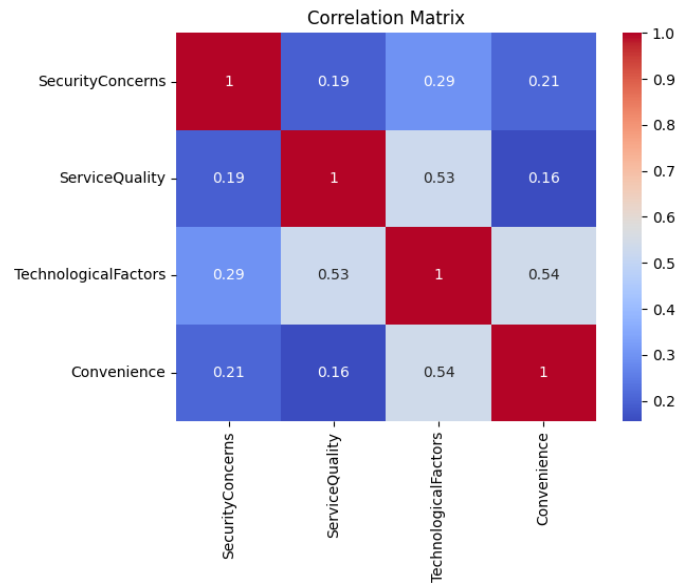

Figure 2: Correlation Matix of the relationship between factors

The correlation matrix in Figure 2 provides insights into the relationships between the factors influencing the usage of mobile banking apps: Security Concerns, Service Quality, Technological Factors, and Convenience. Based on Figure 2, the highest correlation is between Technological Factors and Convenience (0.54). This indicates that respondents that find the usage of mobile banking apps provide them with convenience, are also satisfied with the technological aspects of the mobile banking apps they used. This relationship is important as it highlights that the critical played by the technological aspects of mobile banking apps in enhancing consumers' experience with using mobile banking apps.

Similar correlation between Service Quality and Technological Factors (0.53) which indicates that the technological capabilities of mobile banking apps may also potentially affect the quality of its service. Improving either aspect could potentially enhance the other, leading to better overall user satisfaction. The correlations between Technological Factors with both Service Quality and Convenience may suggest that the factors are interconnected and could jointly influence the usage of mobile banking apps.

**4.5 Spearman Correlation Analysis**

Table 4: Spearman Correlation Analysis

| Variable | Frequency | Spearman Correlation | P-Value |
| --- | --- | --- | --- |
| Security Concerns | Daily | -0.11 | 0.187 |
| Security Concerns | Weekly | 0.15 | 0.067 |
| Security Concerns | Monthly | -0.07 | 0.394 |
| Service Quality | Daily | -0.02 | 0.764 |
| Service Quality | Weekly | 0.04 | 0.598 |
| Service Quality | Monthly | -0.03 | 0.677 |
| Technological Factors | Daily | -0.10 | 0.239 |
| Technological Factors | Weekly | -0.10 | 0.239 |
| Technological Factors | Monthly | -0.10 | 0.239 |
| Convenience | Daily | 0.17 | 0.037 |
| Convenience | Weekly | -0.16 | 0.055 |
| Convenience | Monthly | -0.05 | 0.555 |

The significant positive correlation between Convenience and Daily Frequency (p=value < 0.05) where the p-value is 0.037. shows that there are a significant relationship between perceived convenience and daily usage frequency. the This indicates that the respondents who find the usage of mobile banking apps convenient, they are more likely to use the apps daily. The finding suggest that enhancing the convenience of mobile banking apps could play the role in increasing the daily usage of mobile banking apps among consumers. The correlation between Technological Factors with Daily, Weekly and Monthly Usage appears to be negatively consistent (-0.10), but with a p-value of 0.24, which indicates there no significant relationship between the variables. The consistency in the result suggest a potential weak inverse relationship that could be explored further with a larger sample.

### 4.6 Logistic Regression Analysis

Table 5: Classification Report of Logistic Regression Analysis

| Class | Precision | Recall | F1-score | Support |
| --- | --- | --- | --- | --- |
| Daily | 0.65 | 1.00 | 0.79 | 30 |
| Weekly | 0.00 | 0.00 | 0.00 | 13 |
| Monthly | 0.00 | 0.00 | 0.00 | 3 |
| Accuracy | | | 0.65 | 46 |
| Macro Avg | 0.22 | 0.33 | 0.26 | 46 |
| Weighted Avg | 0.43 | 0.65 | 0.51 | 46 |

The model achieved an accuracy of 0.65, which indicates that the model correctly predicted the frequency of mobile bank apps usage 65% of the time. The precision and recall value for Daily Frequency were both o.65, with an F1-score of 0.79. This indicates that the model achieved a good performance in predicting the Daily Frequency. Meanwhile, the model failed to predict the Weekly and Monthly Frequencies for precision, recall, and F1-scores as all of them valued at 0.00. This reflect the model's poor performance in predicting the Monthly and Weekly Frequencies.

Table 6: Model's Coefficients for predictor variables

| Factor | Daily | Weekly | Monthly |
| --- | --- | --- | --- |
| ServiceQuality | 0.15 | -0.01 | -0.14 |
| SecurityConcerns | -0.09 | 0.43 | -0.34 |
| TechnologicalFactors | -0.13 | 0.32 | -0.19 |
| Convenience | 0.41 | -0.37 | -0.04 |

Table 6 displays the model coefficients of the predictor variables. Based on Table 6, Convenience has a significant positive relationship with Daily Frequency which shows that Convenience is a significant predictor for daily mobile banking apps usage. This result aligns with the previous correlation results. Table 6 also showed that Security Concerns had some influence on weekly usage of mobile banking apps, but Table 5 showed that the model failed to predict Weekly Frequency. The issue occurred likely due to imbalanced data, where the difference of sample size in each category greatly differs.

# 5   Conclusion and Future Work

This study examines the factors that influenced the usage of mobile banking apps among Malaysian consumers. The factors studied were security concerns, technological factors, service quality and convenience. The logistic regression model showed that convenience and security concerns have a significant effect on the usage of mobile banking apps. The model showed that security concerns positively influenced the daily usage of mobile banking apps. The study also found that convenience largely influence the daily usage of mobile banking apps. These findings underscore the importance of addressing security concerns and enhancing consumers' convenience to increase the usage of mobile banking apps among Malaysians.

Future studies could expand on the findings by exploring the relationship between the identified factors and with additional factors such as demographic variables, external economics situations and specific features of mobile banking apps. Longitudinal studies could provide deeper insights into how these factors evolve over time and influence user behavior. Additionally, employing more advanced machine learning models could uncover more complex patterns and relationships within the data, offering a more nuanced understanding of the usage of mobile banking apps. Expanding the scope to include a broader geographic area or diverse consumer segments could also provide a more comprehensive view of the trends and challenges of the usage of mobile banking apps.

c